\documentclass[12pt]{article}

\usepackage{amsmath}        
\usepackage{latexsym}

\newtheorem{teor}{Theorem}[section]

\newtheorem{cor}{Corollary}[section]
\newtheorem{obs}{Remark}[section]
\newtheorem{defin}{Definition}[section]
\newtheorem{exem}{Example}[section]

\newtheorem{algor}{Algorithm}[section]

\newfont{\Mb}{msbm10}
\newcommand{\C}{\mbox{\Mb\symbol{67}}}
\newcommand{\R}{\mbox{\Mb\symbol{82}}}

\begin{document}
\setcounter{equation}{0}
\setcounter{figure}{0}
\setcounter{table}{0}

\hspace\parindent
\thispagestyle{empty}

\bigskip        
\bigskip 
\bigskip
\begin{center}
{\LARGE \bf An Efficient Method for Computing}
\end{center}
\begin{center}
{\LARGE \bf Louvillian First Integrals of}
\end{center}
\begin{center}
{\LARGE \bf Planar Polynomial Vector Fields}
\end{center}

\bigskip

\begin{center}
{\large
$^a$L.G.S. Duarte and $^a$L.A.C.P. da Mota \footnote{E-mails: lgsduarte@gmail.com and lacpdamota@gmail.com}
}

\end{center}

\bigskip
\centerline{\it $^a$ Universidade do Estado do Rio de Janeiro,}
\centerline{\it Instituto de F\'{\i}sica, Depto. de F\'{\i}sica Te\'orica,}
\centerline{\it 20559-900 Rio de Janeiro -- RJ, Brazil}

\bigskip\bigskip
\bigskip
\bigskip

\abstract{Here we present an efficient method to compute Darboux polynomials for polynomial vector fields in the plane. This approach is restricted to polynomial vector fields presenting a Liouvillian first integral (or, equivalently, to rational first order differential equations (rational 1ODEs) presenting a Liouvillian general solution). The key to obtaining this method was to separate the procedure of solving the (non-linear) algebraic system resulting from the equation that translates the condition for the existence of a Darboux polynomial (i.e., from the equation $ D (p) = q \, p $) into feasible steps (procedures that require less memory consumption). We also present a brief performance analysis of the algorithms developed.
}
 
\bigskip
\bigskip
\bigskip
\bigskip
\bigskip
\bigskip



\newpage

\section*{Introduction}
\label{intro}

When we are dealing with the search of Liouvillian first integrals for polynomial vector fields in the plane (or, equivalently, with the search for Liouvillian solutions for rational 1ODEs) the most effective and general methods available belong (probably) to the Darboux-Prelle-Singer (DPS) approach\footnote{For an overview see, for example, \cite{Lli,LlZh}.} \cite{Dar,PrSi,Sin,Man,MaMa,Chr,Nosjpa2002-2,Nosjcam2005,Noscpc2007}. Basically, the central idea behind the DPS approach is the determination of an integrating factor that is composed of certain special polynomials usually called Darboux polynomials (or eigenpolynomials). Thus, the first step on this path is to determine these eigenpolynomials.\footnote{In reality, things are a little more complicated than that because it is necessary to know (in the first place) whether the vector field in question has Darboux polynomials. For research in this field see \cite{CoSc}. Besides, it is also necessary to know if there is a Liouvillian first integral since the algorithms involved are, actually, semi algorithms. Thus, in addition to the developments that directly search for better algorithms to compute Darboux polynomials, there is extensive research that studies the integrability of vector fields presenting Liouvillian first integrals (Darboux integrability). See, for example, \cite{ChLlPaZh,ChGiGiLl,ChLlPaWa1,ChLlPe,ChLlPaWa2,ChLlPaWa3,ChLlPaWa4,Zha}.}

However, this initial step is also the most `difficult': determining Darboux polynomials is, as a rule, computationally costly. Even at low degrees (in practice, around four) the determination of Darboux polynomials through the method of undetermined coefficients (MUC) begins to become impossible in practice. For this reason, there has been, in the last two decades, an extensive search for procedures to improve this situation, that is, to determine the Darboux polynomials in less computational processing time or less memory usage. In other words, a search for more efficient algorithms (faster and/or less costly). In this direction, we highlight the works of G. Chèze, A. Ferragut, A. Gasull, A. Bostan, T. Cluzeau, J.A. Weil, H. Giacomini, T. Combot, C. Christopher, J. Llibre, C. Pantazi, S. Walcher, C. Galindo, F. Monserrat, X. Zhang among others:
\begin{itemize}

\item In \cite{ChLlPaWa1,ChLlPaWa2} C. Christopher, J. Llibre, C. Pantazi and S. Walcher developed an algorithmic approach to determine the planar polynomial vector fields which present a particular Darboux polynomial. Their method can provide an explicit expression for these vector fields if all (finite) singular points of the algebraic curves (defined by the Darboux polynomials) are nondegenerate.

\item In \cite{FeGi}, A. Ferragut and H. Giacomini developed a fast algorithm to compute rational first integrals of a planar polynomial vector field. The algorithm also allowed for the computation of the remarkable curves associated to the rational first integral.

\item In \cite{Che} G. Chèze showed how to compute (for a plane polynomial vector field $D \equiv N\,\partial_x + M\,\partial_y$ of degree $d$) all the irreducible Darboux polynomials of degree smaller than $k$ with ${\cal O}((d\,k\, \log({\cal H}))^{{\cal O}(1)})$ binary operations\footnote{${\cal H} = {\rm max}(\|M\|_\infty,\|N\|_\infty); \,\,\,d = {\rm max}({\rm deg}(M),{\rm deg}(N))$.}. He based his method on the factorization of the extactic curve (a suggestion from J.V. Pereira \cite{ChLlPe}).

\item In \cite{BoChClWe} A. Bostan, G. Chèze, T. Cluzeau and J.A. Weil, based on an idea of Ferragut and Giacomini \cite{FeGi}, could construct algorithms to compute rational ﬁrst integrals via systems of linear equations instead of systems of quadratic equations.

\item In \cite{FeGa} A. Ferragut and A. Gasull construct a method that improves (a lot) the naive method (MUC) for the determination of the Darboux polynomials of a polynomial vector field.

\item In \cite{FeGaMo} A. Ferragut, C. Galindo and F. Monserrat developed a method to compute Darboux first integrals of the type $\prod_i {p_i}^{\alpha_i}$ where the $p_i$ are polynomials defining algebraic curves with only one place at infinity.

\item In \cite{ChCo}, G. Chèze and T. Combot generalize (to the Darbouxian, Liouvillian and Riccati case) the extactic curve introduced by J. Pereira in \cite{Per} and construct new algorithms for computing rational, Darbouxian, Liouvillian or Riccati first integrals of a polynomial planar vector field.

\end{itemize}

In this paper we present a method to determine Darboux polynomials that compose the integrating factors of plane polynomial vector fields (resp. of rational 1ODEs) presenting Liouvillian first integrals (resp. general solutions). It is organized as follows:

In the first section we present some basic concepts involved in the Darboux-Prelle-Singer (DPS) approach. 

In the sequence, we develop some concepts and results that allow for the construction of a method. We use then to propose some procedures that make the DPS approach practical even in cases where the integrating factors have Darboux polynomials of very high degree. We also present examples in order to clarify the procedures.

In the third section we formulate the set of steps of the (possible) algorithms and discuss their performance by making a comparison of memory usage and CPU processing time (we also make comparisons with the method of undetermined coefficients -- MUC).

Finally, we present our conclusions and point some directions for further our work.

\section{Some basic concepts and results in the DPS approach}
\label{dpsapp}

In this section we will present the classical procedures (using the MUC) to find elementary first integrals (Prelle-Singer method) and non elementary Liouvillian first integrals (Christopher-Singer method) for plane polynomial vector fields.

\subsection{The Prelle-Singer (PS) method}
\label{psm} 

Consider a polynomial 2D dynamical system 
\begin{equation}
\label{sd2}
\left\{ \begin{array}{l}
\dot{x} = N(x,y), \\
\dot{y} = M(x,y),
\end{array} \right.
\end{equation}
where $M$ and $N$ are coprime polynomials in $\C[x,y]$. A first integral $I$ of the system (\ref{sd2}) is a function that is constant over the solutions of (\ref{sd2}).

\begin{defin}
Let $L(x,y)$ be a Liouvillian field extention of $\C(x,y)$. A function $I \in L(x,y)$ is said to be a {\bf Liouvillian first integral} of the system if $D(I)=0$, where $D \equiv N\, \partial_x + M\, \partial_y$ is the {\bf vector field associated} (or the {\bf Darboux operator associated}) with the system {\em (\ref{sd2})}.
\end{defin}

\begin{obs}
If  $I \in E(x,y)$, where $E(x,y)$ is an elementary\footnote{For a formal definition of Liouvillian or elementary field extentions see \cite{Dav}.} field extention of $\C(x,y)$, $I$ is said to be an {\bf elementary first integral}.
\end{obs}

\begin{obs}
Since the system {\em (\ref{sd2})} is autonomous, we can divide $\dot{y}$ by $\dot{x}$ to obtain a rational 1ODE
\begin{equation}
\label{r1ode}
y' \equiv \frac{dy}{dx} = \frac{M(x,y)}{N(x,y)}.
\end{equation}
For this rational 1ODE, the function $I$ defines its general solution in the implicit form: $I(x,y) = c$. Besides, the 1-form $\gamma \equiv M\,dx - N\,dy$ is null over the solutions of the 1ODE {\em (\ref{r1ode})} and, therefore, the exact 1-form $\omega \equiv dI$ is proportional to $\gamma$, i.e., $\omega = R\,\gamma$. The function $R$ is commonly called {\bf integrating factor}.
\end{obs}

\medskip

The PS method can be easily understood if we pay attention to the following results obtained by Prelle and Singer \cite{PrSi}:

\begin{teor}[Prelle-Singer]$\!\!:$
\label{ps}
\

\noindent
If the system {\em (\ref{sd2})} presents an elementary first integral then

\noindent
(i) It presents one of the form $I = W_0 + {\sum^K_{j=1}}\, c_j\,\ln(W_j)$, $W$'s are algebraic functions.

\noindent
(ii) The 1-form $M\,dx - N\,dy$ presents an integrating factor of the form $R=\prod_i {p_i}^{n_i}$, where the $p_i$ are irreducible polynomials and the $n_i$ are rational numbers.

\end{teor}

\noindent 
{\bf Proof:} For a proof see \cite{PrSi}.

\medskip

\begin{defin}
Let $p(x,y)\,\in\, \C[x,y]$. the polynomial $p$ is said to be a {\bf Darboux polynomial} of the vector field $D$ if $D(p)=q\,p$, where $D \equiv N\, \partial_x + M\, \partial_y$ and $q$ is a polynomial in $\C[x,y]$ which is called {\bf cofactor} of $p$.
\end{defin}

\medskip

\begin{cor} 
\label{dppmynx}
If the hypothesis of the theorem \ref{ps} is satisfied then
$$
\sum_i n_i \, \frac{D(p_i)}{p_i} = - \mathbf{div}(D),
$$
where $D$ is the vector field associated with {\em (\ref{sd2})}, $p_i$ are Darboux polynomials of $D$ and {\bf div} stands for divergent.
\end{cor}

\noindent
{\bf Proof:} For a proof see section 2 of \cite{Noscpc2002}.

\bigskip

\noindent
The corollary \ref{dppmynx} is the key for the PS method: 

\noindent
{\bf Procedure (sketch):}

\begin{enumerate}

\item Construct the candidates for $p$ and $q$ with undetermined coefficients. Let's call them: $p_c$ and $q_c$. Substitute then in the equation $D(p_c)-q_c\,p_c=0$.

\item  Collect the equation in the variables $(x,y)$ obtaining a set of (quadratic) equations for the coefficients of the candidates. Solve this set of equations to the undetermined coefficients. 

\item Substitute the solutions in the equation $\sum_i n_i \, q_i + N_x + M_y=0$ (see corollary \ref{dppmynx}) and collect the equation in the variables $(x,y)$. Solve the set of equations for the $n_i$.

\item Construct the integrating factor $R=\prod_i {p_i}^{n_i}$ and find (by quadratures) the elementary first integral $I$.

\end{enumerate}

\medskip

\subsection{The Christopher-Singer (CS) method}
\label{csm}

The CS method follows from the results obtained by M. Singer \cite{Sin} and C. Christopher \cite{Chr}:

\begin{teor}[Singer]$\!\!:$
\label{sing}
\

\noindent
If the 1-form $M\,dx - N\,dy$ associated with the system {\em (\ref{sd2})} presents a Liouvillian first integral then it has a integrating factor of the form
\begin{equation}
\label{teosin}
R(x,y) = \exp \left[ \int U(x,y)\,dx + V(x,y)\,dy \right],
\end{equation}
where $U$ and $V$ are rational functions with $U_y=V_x$ so that the line integral {\em (\ref{teosin})} is well defined.

\end{teor}

\noindent 
{\bf Proof:} For a proof see \cite{Sin}.

\medskip

\begin{teor}[Christopher]$\!\!:$
\label{chri}
\

\noindent
If the system {\em (\ref{sd2})}  has an integrating factor of the form {\em (\ref{teosin})} where $U$ and $V$ are rational functions with $U_y=V_x$, then there exists a integrating factor of the system {\em (\ref{sd2})} of the form
\begin{equation}
\label{teochr2}
R(x,y) = \exp \left(\frac{A}{B}\right)\,\prod_i {p_i}^{n_i},
\end{equation}
where $A$, $B$ and the $p_i$ are polynomials in $x$ and $y$. 

\end{teor}

\noindent
{\bf Proof:} For a proof see \cite{Chr,Nosjpa2002-2}.

\begin{defin}
Let $p(x,y)\,\in\, \C[x,y]$. the polynomial $p$ is said to be a {\bf Darboux polynomial} of the vector field $D$ if $D(p)=q\,p$, where $D \equiv N\, \partial_x + M\, \partial_y$ and $q$ is a polynomial in $\C[x,y]$ which is called {\bf cofactor} of $p$.
\end{defin}

\medskip

\begin{cor} 
\label{dppchr}
$B$ and the $p_i$ are Darboux polynomials of the vector field $D$ and $\displaystyle{D\left(\frac{A}{B}\right)}$ is a polynomial.
\end{cor}

\noindent
{\bf Proof:} For a proof see \cite{Chr,Nosjpa2002-2}.

\bigskip

\noindent
The corollary \ref{dppchr} allows for the construct of a semi algorithm in the line of the PS method: 

\noindent
{\bf Procedure (sketch):}

\begin{enumerate}

\item Construct the candidates for $p$ and $q$ with undetermined coefficients. Let's call them: $p_c$ and $q_c$. Substitute then in the equation $D(p_c)-q_c\,p_c=0$.

\item  Collect the equation in the variables $(x,y)$ obtaining a set of (quadratic) equations for the coefficients of the candidates. Solve this set of equations to the undetermined coefficients. 

\item Construct all possible candidates for $B$ (up to a certain degree) and a candidate for $A$. Substitute the solutions) in the equation $B_c\,D(A_c)-A_c\,D(B_c) + {B_c}^2\,(\sum_i n_i \, q_i + N_x + M_y)=0$ (see \cite{Nosjpa2002-2}) and collect the equation in the variables $(x,y)$. Solve the set of equations for the $n_i$ and the coefficients of $A_c$.

\item Construct the integrating factor $R=\exp(A/B)\,\prod_i {p_i}^{n_i}$ and find (by quadratures) the elementary first integral $I$.

\end{enumerate}

\newpage
\section{An efficient method to compute first integrals}
\label{1odesip}

The motivation for the definition of the concepts we will present in this section came from a central idea: to separate the non linear task of computing the Darboux polynomials into linear parts or into simpler non linear ones. Basically, our main interest was that the computer memory did not `blow up', i.e., we've wanted to build an algorithm that `divides' the calculation into sequencial parts such that each one of them could be finished. So, in principle, we were not concerned with the increase or decrease in CPU time, but on other hand, we've wanted that the procedure (as a whole) could be completed. With this motivation in mind, we developed some concepts and managed to get results that allowed us to build alternative procedures to determine Darboux polynomials.

\subsection{1ODEs associated through the integrating factor}
\label{1odesatif}

We start by asking ourselves if the (non linear) complexity in the DPs calculation could be decreased if we knew another polynomial differential operator $D_1$ that `shared' (with the $D$ operator) the Darboux polynomials present in the integrating factor $R$. To this end, we decided that a good candidate to generate the $D_1$ operator would be the rational 1ODE that had $R = C$ ($C$ constant) as a general solution, that is, $D_1$ would be defined by the equation $D_1(R) = 0$. In this way, the Darboux polynomials present in the integrating factor $R$ would also be Darboux polynomials of the $D_1$ operator.\footnote{Note that this will happen because we are dealing with vector fields admitting a Liouvillian first integral. In this case there exists an integrating factor $R$ of the form $\exp \left(\frac{A}{B}\right)\,\prod_i {p_i}^{n_i}$, where $A$, $B$ and the $p_i$ are polynomials in $x$ and $y$ (see Singer theorem \ref{sing}).}

Let's start by defining a rational 1ODE that has the property mentioned above: consider that the rational 1ODE$_0$ 
\begin{equation}
\label{1oder0}
y'=\frac{M_0(x,y)}{N_0(x,y)} = \phi_0(x,y),
\end{equation}
where $M_0$ and $N_0$ are coprime polynomials, presents a Liouvillian general solution $I_0(x,y)=C$. Then, the vector field defined by $D_0 \equiv N_0 \partial_x + M_0 \partial_y$ (the {\em Darboux operator} associated with the 1ODE$_0$) presents and an integrating factor $R_0$ of the form 
\begin{equation}
\label{r0}
R_0= {\rm e}^{Z_0}\,\prod_i {p_i}^{n_i},
\end{equation}
where $Z_0=A/B$ is a rational function ($A$ and $B$ are coprime polynomials), $p_i$ are irreductible polynomials and $n_i$ are rational numbers. Besides, the polynomial $B$ and the polynomials $p_i$ are Darboux polynomials of the operator $D_0$.\footnote{See \cite{Sin,Chr,Nosjpa2002-2}.}

\begin{defin}
Consider the rational 1ODE$_0$ {\em (\ref{1oder0})}. We call {\bf 1ODE associated through the integrating factor} an 1ODE  (denoted by 1ODE$_1$)
\begin{equation}
\label{1odeatif}
y'=\frac{M_1(x,y)}{N_1(x,y)} = \phi_1(x,y),
\end{equation}
whose associated vector field ($D_1 \equiv N_1 \partial_x + M_1 \partial_y$) obeys the relation $D_1(R_0)=0$.
\end{defin}

\begin{obs}
\label{classd1}
Since any function of the first integral $I_0$ (which is invariant under the action of the vector field $D_0$) multiplied by the integrating factor $R_0$ is itself an integrating factor, the equation $D_1(R_0)=0$ defines an equivalence class of vector fields: $[D_1]$.
\end{obs}

In order to work with only one associated 1ODE, when we mention 1ODE$_1$ we will be referring to the 1ODE that represents the class $[D_1]$ and, since the equivalence class is defined by the form of the integrating factor $R_0$, to choose a representative of the class, we only have to choose (for a given 1ODE$_0$) a `canonical' integrating factor. We will do this as follows:
\begin{itemize}
\item $I_0$ is a non elementary Liouvillian function $\,\rightarrow\,$ $R_0$ is a Darboux function (${\rm e}^{Z_0}\, \prod_i {p_i}^{n_i}$) or an algebraic function ($\prod_i {p_i}^{n_i}$ where at least one of the $n_i$ is a non integer rational number).
\item $I_0$ is a non algebraic elementary function $\,\rightarrow\,$ $R_0$ is an algebraic function ($\prod_i {p_i}^{n_i}$) or a rational function.
\item $I_0$ is a rational function $\,\rightarrow\,$ $R_0$ is the rational function with least degree such that $R_1/R_0 = 1/\prod_i {p_i}$.
\end{itemize}

Once we have defined the 1ODE$_1$ associated with the 1ODE$_0$ (\ref{1oder0}), we can enunciate the following result:

\begin{teor}
\label{pepd}
Consider the 1ODE$_0$ {\em (\ref{1oder0})} defined as above. Then the Darboux polynomials of $D_0$ present in the integrating factor $R_0$ are also Darboux polynomials (eigenpolynomials) of the operator $D_1$.
\end{teor}

\noindent
{\bf Proof:} From the hypotheses we have that ${D_1}(R_0)=0$. Therefore,
\begin{equation}
\label{proofpepd1}
{D_1}(R_0)={D_1}\left({\rm e}^{Z_0}\,\prod_i {p_i}^{n_i} \right)=D_1\left( \frac{A}{B}\right) + \sum_i n_i \frac{{D_1}(p_i)}{p_i}=0.
\end{equation}
Since the $p_i$ are irreductible polynomials, the conclusion follows easily from theorem 1 of \cite{Nosjpa2002-2}. $\,\,\,\Box$

\begin{teor}
\label{overphipde}
Consider the 1ODE$_0$ and the 1ODE$_1$ defined as above. Then the function $\phi_1$ obeys the following 1PDE:
\begin{eqnarray}
&&
M_0\left({M_0}_{y}+{N_0}_{x} \right) {\phi_1}_{y}+ N_0\left( {M_0}_{y}+{N_0}_{x} \right) {\phi_1}_{x}+
\nonumber \\ [2mm]
&&
+\left( {M_0}_{y}\,{N_0}_{y}-{M_0}_{yy}\,{N_0}-{N_0}{N_0}_{xy}+{N_0}_{x}\,{N_0}_{y} \right) {{\phi_1}}^{2}+ 
\nonumber \\ [2mm]
&&
+\left( {M_0}{M_0}_{yy}+{M_0}{N_0}_{xy}-{M_0}_{xy}\,{N_0}-{{M_0}_{y}}^{2}-{N_0}{N_0}_{xx}+{{N_0}_{x}}^{2} \right) {\phi_1}+
\nonumber \\ [2mm]
&&
+ {M_0}{M_0}_{xy}+{M_0}{N_0}_{xx}-{M_0}_{x}\,{M_0}_{y}-{M_0}_{x}\,{N_0}_{x}=0.
\label{pdephi}
\end{eqnarray}
\end{teor}

\noindent
{\bf Proof:} From the hypothesis, $D_1(R_0)=0$. So, 
\begin{equation}
\label{eqt2p1}
N_1\,{R_0}_x+M_1\,{R_0}_y=0 \,\,\,\Rightarrow\,\,\,{R_0}_x+\frac{M_1}{N_1}\,{R_0}_y=0\,\,\,\Rightarrow\,\,\, \frac{{R_0}_x}{R_0}+\phi_1\,\frac{{R_0}_y}{R_0}=0.
\end{equation}
Besides, $R_0$ obeys the equation $D_0(R_0)= - R_0\,({N_0}_x+{M_0}_y)$ and so:
\begin{equation}
\label{eqt2p2}
N_0\,{R_0}_x+M_0\,{R_0}_y=-R_0\,({N_0}_x+{M_0}_y) \,\,\,\Rightarrow\,\,\,N_0\,\frac{{R_0}_x}{R_0}+M_0\,\frac{{R_0}_y}{R_0}=-R_0\,({N_0}_x+{M_0}_y).
\end{equation}
We have that $\frac{{R_0}_x}{R_0}=\partial_x(\ln(R_0)),\,\frac{{R_0}_y}{R_0}=\partial_y(\ln(R_0))$ and, solving the equations (\ref{eqt2p1}) and (\ref{eqt2p2}) for them we get
\begin{eqnarray}
\label{eqt2p3}
\partial_x(\ln(R_0)) &=& \frac{\phi_1\,({N_0}_x+{M_0}_y)}{M_0-\phi_1\,N_0},  \\ [2mm]
\partial_y(\ln(R_0)) &=& \frac{-({N_0}_x+{M_0}_y)}{M_0-\phi_1\,N_0}. \label{eqt2p4}
\end{eqnarray}
The equality $\partial_y(\partial_x(\ln(R_0)))=\partial_x(\partial_y(\ln(R_0)))$ implies that
\begin{equation}
\label{eqt2p5}
\partial_y\left(\frac{\phi_1\,({N_0}_x+{M_0}_y)}{M_0-\phi_1\,N_0}\right)+\partial_x\left(\frac{({N_0}_x+{M_0}_y)}{M_0-\phi_1\,N_0}\right)=0,
\end{equation}
and after a straightforward calculation we arrived at the 1PDE (\ref{pdephi}). $\,\,\,\Box$

\bigskip

\subsection{`Linear' computation of cofactors}
\label{lncc}

The idea in this section is to show how we can construct a procedure divided into two stages: The first will compute the cofactors and then second will determine the Darboux polynomials. Note that if we had, at our disposal, a linear procedure to calculate the cofactors before the Darboux polynomials, we would be able to split the determining of the coefficients of the Darboux polynomials and their respective cofactors (a calculation involving second-degree algebraic systems resulting from the condition $D(p) = q\,p$) in two very simple procedures (since, knowing the cofactor $q$ {\em a priori}, the equation for the coefficients of $p$ obtained from the condition $D(p) = q\,p$ would be linear). To achieve this we made use of the fact that the Darboux polynomials present in the integrating factor $R_0$ are also Darboux polynomials of the operator $D_1$. Based on this knowledge we can find a linear relation involving the cofactors of Darboux polynomials $p_i$ of $D_0$ and $D_1$. 

\begin{defin}
Consider the 1ODE$_0$ and the 1ODE$_1$ defined as above and let $p$ be a Darboux polynomial of the operators $D_0$ and $D_1$ with cofactors given by, respectively, $q_0$ and $q_1$. The operator $\overline{D}$ will be defined as
\begin{equation}
\label{defdb}
\overline{D} \equiv q_1\,D_0 - q_0\,D_1
\end{equation}
\end{defin}

\medskip

\begin{obs}
Note that
$
\overline{D}(p) = q_1\,D_0(p) - q_0\,D_1(p) = q_1\,q_0\,p - q_0\,q_1\,p = 0.
$
\end{obs}

\medskip

\noindent
We can also note that the commutator of the operator $\overline{D}$ with $D_0$ would also have $p$ as a Darboux polynomial. We can express the commutator $\left[\overline{D},D_0\right]$ as (directly from the definition of $\overline{D}$)
\begin{eqnarray}
\left[\overline{D},D_0\right] &=& q_1{D_0}^2-q_0D_1D_0-D_0(q_1)\,D_0 + D_0(q_0)\,D_1-q_1{D_0}^2+q_0D_0D_1 =
\nonumber \\ [2mm]
&=& - D_0(q_1)\,D_0 + D_0(q_0)\,D_1  + q_0 \left[D_0,D_1\right], \label{comd10d}
\end{eqnarray}

\noindent
where the commutator $\left[D_0,D_1\right]$ is given by

\begin{equation}
\label{comd0d1}
\left[D_0,D_1\right] =  F_0 D_0 + F_1 D_1
\end{equation}
where $F_0$ and $F_1$ are given by      
\begin{equation}
\label{defF0}
F_0 \equiv \frac{{N_1}(D_0(M_1))-D_1(M_0))-{M_1}(D_0(N_1))-D_1(N_0))}{\Delta},
\end{equation}
\begin{equation}
\label{defF1}
F_1 \equiv  \frac{{M_0}(D_0(N_1))-D_1(N_0))-{N_0}(D_0(M_1))-D_1(M_0))}{\Delta},
\end{equation}
and $\Delta \equiv M_0\,N_1-M_1\,N_0$.
\bigskip

We can now show the result mentioned above. It can be enunciated as:

\begin{teor}
\label{oversuperes}
Consider the 1ODE$_0$ and the 1ODE$_1$ defined as above. If $p$ is a Darboux polynomial of the operators $D_0$ and 
$D_1$ (as defined above), i.e., if 
\begin{eqnarray}
D_0(p)&=&q_0\,p,
\nonumber \\ [2mm]
D_1(p)&=&q_1\,p,
\nonumber 
\end{eqnarray}
where $q_0$ and $q_1$ are the cofactors of $p$, then
\begin{equation}
\label{oversuperes2}
D_0(q_1)-D_1(q_0)=q_0\,F_0+q_1\,F_1,
\end{equation}
where $F_0$ and $F_1$ are given by {\em (\ref{defF0})} and {\em (\ref{defF1})}, respectively.
\end{teor}

\noindent
{\bf Proof:} From the definition of the operator $\overline{D}$ we have that 
\begin{equation}
\label{eqpro1}
\left[\overline{D},D_0\right] (p) = \overline{D}\underbrace{\left(D_0(p)\right)}_{q_0p} - D_0\underbrace{\left(\overline{D}(p)\right)}_{0} = p\, \overline{D}\left(q_0\right) = p\,\left( q_1\,D_0\left(q_0\right) - q_0\,D_1\left(q_0\right)\right).
\end{equation}

\noindent
From equation (\ref{comd10d}) we have that 
$$
\left[\overline{D},D_0\right] (p) = - D_0(q_1)\,\underbrace{D_0(p)}_{q_0\,p} + D_0(q_0)\,\underbrace{D_1(p)}_{q_1\,p}  + q_0 \underbrace{\left[D_0,D_1\right](p)}_{ F_0 D_0(p) + F_1 D_1(p)} =
$$

\begin{equation}
\label{eqpro2}
\,\,\,\,\,\,\,\,\,\,\,\,\,\,\,\,\,\,\,\,\,\,\,\,\,\, = p\,\left( - D_0(q_1)\,{q_0} + D_0(q_0)\,{q_1}  + q_0 \left(q_0\, F_0  + q_1\, F_1 \right)\right).
\end{equation}
So, since $p$ is a commom factor of equations (\ref{eqpro1}) and (\ref{eqpro2}), we have: 
\begin{equation}
\label{eqpro3}
q_1\,D_0\left(q_0\right) - q_0\,D_1\left(q_0\right) = - D_0(q_1)\,{q_0} + D_0(q_0)\,{q_1}  + q_0 \left(q_0\, F_0  + q_1\, F_1 \right).
\end{equation}
Eliminating the term $D_0(q_0)\,{q_1}$ that appears on both sides of equation (\ref{eqpro3}), we see that all the remaining terms have the factor $q_0$. So, we can write equation (\ref{eqpro3}) as $D_0(q_1)-D_1(q_0)=q_0\,F_0+q_1\,F_1$. $\,\,\,\Box$

\begin{obs}
Note that the equation {\em (\ref{oversuperes2})} is linear in the cofactors $q_0$ and $q_1$ and, therefore, on the coefficients of its monomials. So, if we could obtain the cofactors $q_0$ and $q_1$ from equation {\em (\ref{oversuperes2})}, we would be able to use the classical equation $D_0(p)=q_0\,p$ to obtain the coefficients of $p$ in a linear fashion.
\end{obs}

\noindent
Let's see this part of the process in an example.

\begin{exem}
\label{examp1}
Consider the rational 1ODE$_0$ given by 
\begin{equation}
\label{examp1eq1}
\phi_0={\frac {-{y}^{7}+x{y}^{4}-{x}^{2}y+{y}^{2}}{2\,x{y}^{6}-7\,{x}^{2}{y}^
{3}+2\,{x}^{3}+3\,xy}}.
\end{equation}
Knowing that $\phi_1$ is given by 
\begin{equation}
\label{examp1eq2}
\phi_1={\frac {{y}^{6}-3\,x{y}^{3}+2\,y}{-7\,x{y}^{5}+{x}^{2}{y}^{2}+5\,{y}^{
3}+x}},
\end{equation}
let's use equation {\em (\ref{oversuperes2})} to compute the cofactors and (only) then determine the Darboux polynomials.
\end{exem}

\noindent
{\bf Procedure:}

\begin{enumerate}

\item The operators $D_0$ and $D_1$ are given by:
\begin{eqnarray}
D_0 &=& (2\,x{y}^{6}-7\,{x}^{2}{y}^{3}+2\,{x}^{3}+3\,xy)\,\partial_x + (-{y}^{7}+x{y}^{4}-{x}^{2}y+{y}^{2})\,\partial_y,
\nonumber \\ [2mm]
D_1&=& (-7\,x{y}^{5}+{x}^{2}{y}^{2}+5\,{y}^{3}+x)\,\partial_x + ({y}^{6}-3\,x{y}^{3}+2\,y)\,\partial_y.
\nonumber 
\end{eqnarray}

\item $F_0$ and $F_1$ are given by:
\begin{eqnarray}
F_0 &=& -{\frac {-6\,{y}^{11}+26\,x{y}^{8}+2\,{x}^{2}{y}^{5}-21\,{y}^{6}+2\,{x
}^{3}{y}^{2}-7\,x{y}^{3}+2\,{x}^{2}+2\,y}{-{y}^{6}-2\,x{y}^{3}+{x}^{2}
+y}},
\nonumber \\ [2mm]
F_1&=& 5\,{\frac {{y}^{4} \left( {y}^{8}-2\,x{y}^{5}+3\,{x}^{2}{y}^{2}-{y}^{3
}-x \right) }{-{y}^{6}-2\,x{y}^{3}+{x}^{2}+y}}.
\nonumber 
\end{eqnarray}

\item The degree of $M_0$ is 7 and the degree $N_0$ is 7 too. So, the maximum degree for the cofactors (of $D_0$) is 6. On the other hand, the degree of $M_1$ is 6 and this is also the degree of $N_1$. So, the maximum degree for the cofactors (of $D_1$) is 5. We can construct the cofactor candidates $Q_0$ and $Q_1$ as:
\begin{eqnarray}
Q_0 &=& a_{0}+a_{18}\,{x}^{2}y+a_{19}\,{x}^{2}{y}^{2}+a_{20}\,{x}^{2}{y}^{4}+a_{21}\,{x}^{3}y+a_{22}\,{x}^{3}{y}^{2}+
a_{23}\,{x}^{3}{y}^{3}+
\nonumber \\ [2mm]
&& 
+a_{24}\,{x}^{4}y\!+\!a_{25}\,{x}^{5}y\!+a_{26}\,{y}^{2}{x}^{4}\!+a_{27}\,{y}^{3}{x}^{2}\!+a_{13}\,xy\!+\!a_{14}\,x{y}^{2}\!+\!a_{15}\,x{y}^{3}+
\nonumber \\ [2mm]
&& 
+a_{16}\,x{y}^{4}+a_{17}\,x{y}^{5}+a_{9}\,{y}^{3}+a_{10}\,{y}^{4}+a_{11}\,{y}^{5}+a_{12}\,{y}^{6}+a_{1}\,x+a_{2}\,y+
\nonumber \\ [2mm]
&&
+a_{3}\,{x}^{2}+a_{4}\,{x}^{3}+a_{5}\,{x}^{4}+a_{6}\,{x}^{5}+a_{7}\,{x}^{6}+a_{8}\,{y}^{2},
\nonumber \\ [5mm]
Q_1&=& b_{0}\!+b_{1}\,x\!+b_{2}\,y\!+b_{3}\,{x}^{2}\!+b_{4}\,{x}^{3}\!+\!b_{5}\,{x}^{4}+b_{6}\,{x}^{5}+b_{7}\,{y}^{2}+
b_{8}\,{y}^{3}+b_{9}\,{y}^{4}+
\nonumber \\ [2mm]
&&
+b_{10}\,{y}^{5}+b_{11}\,xy+b_{12}\,x{y}^{2}+b_{13}\,x{y}^{3}+b_{14}\,x{y}^{4}+b_{15}\,{x}^{2}y+b_{16}\,{x}^{2}{y}^{2}+
\nonumber \\ [2mm]
&&
+b_{17}\,{x}^{3}y+b_{18}\,{x}^{3}{y}^{2}+b_{19}\,{x}^{4}y+b_{20}\,{y}^{3}{x}^{2}.
\nonumber 
\end{eqnarray}

\item We use the equation $D_0(Q_1)-D_1(Q_0)-Q_0\,F_0-Q_1\,F_1=0$ and collect the numerator in the variables $(x,y)$. We obtain a set of (linear) equations for the coefficients of the candidates $Q_0$ and $Q_1$: (we present only a few to exemplify)
\begin{eqnarray}
\left\{ a_{1},2\,a_{0},-41\,a_{6},-39\,a_{6},-3\,
a_{6},47\,a_{6},-52\,a_{7},-48\,a_{7},-4\,a_{7},56\,a_{7},7\,b_{19},-55\,b_{6},
\right.
\nonumber \\ [2mm]
10\,b_{6},-a_{11}-b_{9},-9\,a_{16}-4\,b_{13},-16\,a_{20}-6\,b_{20}-24\,a_{23}-9\,b_{18},-40\,a_{25}-15\,b_{6},
\nonumber \\ [2mm]
\left. \cdots,-20\,a_{5}+8\,a_{8}-7\,a_{23}+5\,b_{13}-21\,b_{3}-8\,a_{18}-7\,a_{16}+3\,b_{2}+11\,b_{18} \right\}.\,\,\,\,\,\,\,\,\,\,\,\,\,\,\,\,\,
\nonumber 
\end{eqnarray}

\item We solve the set of equations for the coefficients of the candidates $Q_0$ and $Q_1$, obtaining the solution
\begin{eqnarray}
 \left\{ a_{0}=0,a_{1}=0,a_{10}=0,a_{11}\!=0,a_{12}=a_{12},a_{13}=0,a_{14}=0,a_{15}=a_{15},a_{16}=0,\,
\right.
\nonumber \\ [2mm]
a_{17}\!=0,a_{18}\!=0,a_{19}=0,a_{2}=-a_{12},a_{20}=0,a_{21}=0,a_{22}=0,a_{23}=0,a_{24}\!=0, \,\,\,
\nonumber \\ [2mm]
a_{25}\!=0,a_{26}\!=0,a_{27}\!=0,a_{3}\!=a_{3},a_{4}=0,a_{5}=0,a_{6}=0,a_{7}=0,a_{8}=0,a_{9}\!=0,\,\,\,
\nonumber \\ [2mm]
b_{0}\!=\!-a_{3}-\!a_{12},b_{1}\!=0,b_{10}\!=a_{15}-\!a_{12}+\!a_{3},b_{11}\!=\!0,b_{12}\!=a_{15}+\!a_{12}+\!3\,a_{3},b_{2}\!=0,\,\,
\nonumber \\ [2mm]
b_{13}\!=\!0,b_{14}\!=\!0,b_{15}\!=0,b_{16}\!=0,b_{17}=0,b_{18}=0,b_{19}=0,b_{20}=0,b_{3}=0,b_{4}=0,\,\,
\nonumber \\ [2mm]
\left.
b_{5}=0,b_{6}=0,b_{7}=0,b_{8}=0,b_{9}=0 \right\} \qquad \qquad \qquad \qquad \qquad \qquad \qquad \qquad \qquad 
\nonumber 
\end{eqnarray}

\noindent
that leads to the possible candidates for the cofactors:
\begin{eqnarray}
Q_0 &=& \left( {y}^{6}-y \right) a_{12}+x{y}^{3}\,a_{15}+{x}^{2}\,a_{3},
\nonumber \\ [2mm]
Q_1&=& \left( -{y}^{5}+x{y}^{2}-1 \right) a_{12}+ \left( {y}^{5}+x{y}^{2} \right) a_{15}+ \left( {y}^{5}+3\,x{y}^{2}-1 \right) a_{3}.
\nonumber 
\end{eqnarray}

\item We can construct a Darboux polynomial candidate $P$ (of a chosen degree) in the same way we did for the cofactors candidates. Then, we pick one of the equations $D_0(P)-Q_0\,P=0$ or $D_1(P)-Q_1\,P=0$ and collect it in the variables $(x,y)$. For the degrees 1 and 2 the only Darboux polynomial obtained is $y$. For the degree 3 we have:
\begin{equation}
P = p_{4}\,{x}^{3}+p_{6}\,{y}^{3}+p_{8}\,x{y}^{2}+p_{9
}\,{x}^{2}y+p_{3}\,{x}^{2}+p_{5}\,{y}^{2}+p_{7}\,xy+p_{1}\,x+p_{2}\,y+p_{0}.
\nonumber 
\end{equation}
Using the equation $D_1(P)-Q_1\,P=0$ and proceeding in a manner analogous to the preceding steps we obtain the following solutions:
\begin{eqnarray}
 \left\{ a_{12}=-a_{3}-1,a_{15}=-2\,a_{3}+2,a_{3}=a_{3},p_{0}=0,p_{1}=-p_{6},p_{2}=0,p_{3}=0, 
\right. 
\nonumber \\ [2mm]
\left. p_{4}=0,p_{5}=0,p_{6}=p_{6},p_{7}=0,p_{8}=0,p_{9}=0 \right\} , \qquad\qquad\qquad\qquad\qquad\qquad
\nonumber \\ [2mm]
\left\{ a_{12}\!=\!-a_{3},a_{15}\!=\!-2\,a_{3}-5,a_{3}=a_{3},p_{0}=p_{0},p_{1}=0,p_{2}=0,p_{3}=0,p_{4}=0,
\right. 
\nonumber \\ [2mm]
\left. p_{5}=0,p_{6}=0,p_{7}=0,p_{8}=-p_{0},p_{9}=0 \right\} , \qquad\qquad\qquad\qquad\qquad\qquad\qquad\,\,
\nonumber \\ [2mm]
\left\{ a_{12}\!=\!-a_{3}-2,a_{15}\!=\!-2\,a_{3}-1,a_{3}\!=\!a_{3},p_{0}\!=\!0,p_{1}\!=\!0,p_{2}\!=\!p_{2},p_{3}\!=0,p_{4}\!=0,
\right. 
\nonumber \\ [2mm]
\left. p_{5}=0,p_{6}=0,p_{7}=0,p_{8}=0,p_{9}=0 \right\}. \qquad\qquad\qquad\qquad\qquad\qquad\qquad\quad\,\,\,
\nonumber 
\end{eqnarray}
leading, respectively, to the following Darboux polynomials:

$-p_{6}\, \left( -{y}^{3}+x \right),$ $-p_{0}\, \left( x{y}^{2}-1 \right)$ and $p_{2}\,y$.

\end{enumerate}

\medskip

\begin{obs}
\label{obsexamp1}
Some comments:

\begin{itemize}

\item The Darboux polynomials obtained in the step 6 are sufficient to determine an integrating factor. We just need to use the final part of the PS method described in section \ref{psm}. By doing this we obtain 
$${\frac {y}{ \left( x{y}^{2}-1 \right)  \left( -{y}^{3}+x \right) ^{2}}}$$
as an integrating factor and 
$${\frac {x}{-{y}^{3}+x}}-\ln  \left( x{y}^{2}-1 \right)$$
as a first integral for the vector field $D_0$.

\item The Darboux polynomials $-{y}^{3}+x$ and $x{y}^{2}-1$ are of third degree and they are very difficult to be obtained by the naive method (MUC). We will discuss the CPU time and the use of computer memory involved in the processes in a later section.

\item The definition of the $\overline{D}$ operator involves the cofactors of a Darboux polynomial common to the $D_0$ and $D_1$ operators. Since we use only this premise in the deduction of the equation {\em (\ref{oversuperes2})}, it must apply to cofactors of any Darboux polynomial common to the operators $D_0$ and $D_1$. Thus, if the number of Darboux polynomials (in common) is greater than 1, we must have some uncertainty in the solution of the linear equation {\em (\ref{oversuperes2})}.

\item Once the cofactors could not be determined precisely by the equation {\em (\ref{oversuperes2})}, the computation of the Darboux polynomials was not linear in view of the remaining undetermined $p_{6},\, p_{0}$ and $p_{2}$.

\item We use only the equation $D_1(P)-Q_1\,P=0$. What would happen if we used the equation $D_0(P)-Q_0\,P=0$? Would the difficulty be the same in `more complicated' cases? What if we used both sets of equations for the coefficients of the Darboux polynomial? 

\item Although CPU time and memory usage costs are minimal in this example, they represent the cost of only part of the algorithm. For the method to be complete, it is necessary to include the determination of the polynomials $M_1$ and $N_1$. When we tried to use the 1PDE {\em (\ref{pdephi})} to determine them, the result was not encouraging:  the computer memory `overflowed'.
\end{itemize}
\end{obs}

\medskip

\subsection{Obtaining the 1ODE associated}
\label{o1odea}

In the previous section we have shown a method to calculate the Darboux polynomials in two steps: first, one computes the cofactors and, in a later procedure, determines the DPs. However, to apply these procedures we need to know {\em a priori} the vector field $D_1$ which is the Darboux vector field associated with the 1ODE$_1$ (\ref{1odeatif}). Unfortunately, the process of computing the $\phi_1$ using the 1PDE (\ref{pdephi}) is also a non linear (quadratic) one. Besides, it does not seem to be much more efficient than the naive method (MUC) applied to the classical problem, i.e., $D_0(p)=q_0\,p$ (see section \ref{perf}). In this section we show how we can `break up' the computation of $M_1$ and $N_1$ into two steps in an analogous way to what we have done to calculate the Darboux polynomials.

\subsubsection{Two important results}
\label{tir}

We were able to achieve the two-step procedure (that we have mentioned in the end of the previous section) by noting that the definition of the 1ODE$_1$ and the structure of the function $F_0$ imply two results that, used together, allow the elaboration of a more efficient method than the use of 1PDE (\ref{pdephi}) for the computation of the polynomials $M_1$ and $N_1$. We will enunciate the first of these results as follows:

\begin{teor}
\label{delit0}
Consider the 1ODE$_0$ {\em (\ref{1oder0})} and the 1ODE$_1$ {\em (\ref{1odeatif})} and the operators $D_0$ and $D_1$ (as defined above). If the commutator of the the operators $D_0$ and $D_1$ is writen as in {\em (\ref{comd0d1})}, 
\begin{equation}
\left[D_0,D_1\right] =  F_0 D_0 + F_1 D_1, \nonumber
\end{equation}
then at least one of the following afirmatives holds:

\vspace{2mm}

i) $F_0$ is a polynomial; 

\vspace{2mm}

ii) $({M_0}_y+{N_0}_x)$ has a proper polynomial factor in commom with $\Delta$;

\vspace{2mm}

iii) $({M_0}_y+{N_0}_x) | \Delta$;

\vspace{2mm}

\noindent
where $\Delta \equiv M_0\,N_1-M_1\,N_0$.
\end{teor}

\noindent
{\bf Proof:} Applying the operator $\left[D_0,D_1\right]$ to the integrating factor $R_0$ (of the 1ODE$_0$) we have (Remembering that $D_1(R_0)=0$):
\begin{eqnarray}
\left[D_0,D_1\right](R_0) &=& D_0\,\left(D_1(R_0)\right) - D_1\left(D_0(R_0)\right)= - D_1\left(R_0\,({M_0}_y+{N_0}_x)\right)= \nonumber \\ [2mm]
&=& -R_0\, D_1\left({M_0}_y+{N_0}_x\right).
\nonumber
\end{eqnarray}

\noindent
By using $\left[D_0,D_1\right] =  F_0 D_0 + F_1 D_1$ we have that 
$$
\left[D_0,D_1\right](R_0) = F_0 D_0(R_0) + F_1 D_1(R_0)= - F_0\,R_0\,({M_0}_y+{N_0}_x)
$$
and so $D_1\left({M_0}_y+{N_0}_x\right)= - F_0\,({M_0}_y+{N_0}_x)$. Since $D_1\left({M_0}_y+{N_0}_x\right)$ is a polynomial this implies that $F_0\,({M_0}_y+{N_0}_x)$ is also a polynomial. From the hypotheses $F_0$ is given by
$$
F_0 = \frac{{N_1}(D_0(M_1)-D_1(M_0))-{M_1}(D_0(N_1)-D_1(N_0))}{\Delta}.
$$
Defining $P_{N\!F_0} \equiv {N_1}(D_0(M_1)-D_1(M_0))-{M_1}(D_0(N_1)-D_1(N_0))$ and $T_0 \equiv {M_0}_y+{N_0}_x$ in order to simplify the notation, 
we have that $\Delta | (P_{N\!F_0} \, T_0)$. Since $P_{N\!F_0}$ is a polynomial (as well as  $\Delta$ and $T_0$) the conclusion of the theorem follows straightforward: 

\vspace{2mm}
\noindent
$(i)$  If $\Delta | P_{N\!F_0}$ $\Rightarrow$ $F_0$ is polynomial.

\vspace{2mm}
\noindent
$(ii)$ and $(iii)$  If $F_0$ is non polynomial $\Rightarrow$ a factor of $\Delta$ must divide $T_0$ or be equal to it. $\,\,\,\Box$

\bigskip

The second result is a consequence of a deeper investigation into the relationship between $\Delta$ and $T_0$:

\begin{teor}
Consider the 1ODE$_0$ {\em (\ref{1oder0})} and the 1ODE$_1$ {\em (\ref{1odeatif})} and let ${\cal I}$ be defined as
\begin{equation}
\label{defcali}
{\cal I} \equiv \frac{M_0\,N_1-M_1\,N_0}{{M_0}_y+{N_0}_x}.
\end{equation}
Then ${\cal I}$ is an inverse integrating factor for the vector field $D_1$ defined as above.
\end{teor}

\noindent
{\bf Proof:} By hypothesis, we have that $D_1(R_0)=0$ (i.e., $R_0$ is a first integral of $D_1$). Therefore
\begin{eqnarray}
{R_0}_x &=& R_1\,M_1, \nonumber \\ [2mm]
{R_0}_y &=& - R_1\,N_1, \nonumber
\end{eqnarray}
and, so, we can write $M_0\,N_1-M_1\,N_0$ as
$$
M_0\,N_1-M_1\,N_0 = M_0\,\frac{-{R_0}_y}{R_1} -  N_0\,\frac{{R_0}_x}{R_1} = - \frac{D_0(R_0)}{R_1} = \frac{R_0}{R_1} \, ({M_0}_y+{N_0}_x)
$$
Implying that ${\cal I}= \frac{R_0}{R_1}$. Applying $D_1$ we have
$$
D_1({\cal I}) = D_1\left(\frac{R_0}{R_1}\right) = \frac{R_1\,D_1(R_0)-R_0\,D_1(R_1)}{{R_1}^2} = - \frac{R_0}{R_1} \, \frac{D_1(R_1)}{R_1} =  \frac{R_0}{R_1}\, ({M_1}_y+{N_1}_x).
$$
Therefore,
$$\frac{D_1({\cal I})}{{\cal I}} = {M_1}_y+{N_1}_x. \,\,\,\,\,\,\,\,\,\,\,\,\Box$$

\bigskip

These two results allow the development of a strategy to determine $M_1$ and $N_1$ in a more efficient way than the use of 1PDE (\ref{pdephi}) in a large number of cases. 

\medskip

\subsubsection{The two-step procedure}
\label{tsp}

In the previous section we could deduce two results both related to the connection between the polynomial functions $T_0 \equiv {M_0}_y+{N_0}_x$ and $\Delta \equiv M_0\,N_1-M_1\,N_0$. We can express this relationship through the following equation:
\begin{equation}
\label{reldeltat0}
M_0\,N_1-M_1\,N_0 = {\cal I} \, ({M_0}_y+{N_0}_x).
\end{equation}
Applying $D_1$ to (\ref{reldeltat0}) we have (since ${\cal I}$ is an inverse integrating factor for the vector field $D_1$) that
\begin{equation}
\label{d1reldeltat0}
D_1(\Delta) = {\cal I} \, D_1(T_0) + T_0 \, D_1({\cal I}) = {\cal I} \, (D_1(T_0) + T_0 \, T_1),
\end{equation}
where $T_1 \equiv {M_1}_y+{N_1}_x$.
We can use the equations (\ref{reldeltat0}) and (\ref{d1reldeltat0}) to build a method for the case $(iii)$ of theorem \ref{delit0} (the case $T_0 | \Delta$), because whenever this situation occurs the inverse integrating factor ${\cal I}$ is a polynomial. 

\medskip

\noindent
{\bf Procedure (sketch):}

\begin{enumerate}

\item Construct the candidates for $M_1$, $N_1$ and ${\cal I}$ with undetermined coefficients. Let's call them: $M_c$, $N_c$ and ${\cal I}_c$. Substitute then in the equation $\Delta - {\cal I}\, T_0=0$.

\item  Collect the equation in the variables $(x,y)$ obtaining a set of (linear) equations for the coefficients of the candidates. Solve this set of equations to the undetermined coefficients. 

\item Substitute the solution in the equation $D_1(\Delta) - {\cal I} \, (D_1(T_0) + T_0 \, T_1)=0$ and collect the equation in the variables $(x,y)$. Solve the set of equations for the remaining undetermined coefficients.

\item Substitute the solution in the candidates $M_c$ and $N_c$.

\end{enumerate}

\medskip

\noindent
Let's see this procedure in practice:

\begin{exem}
\label{examp2}
Consider the rational 1ODE$_0$ given by 
\begin{equation}
\label{examp2eq1}
\phi_0={\frac {-{y}^{7}+x{y}^{4}-{x}^{2}y+{y}^{2}}{2\,x{y}^{6}-7\,{x}^{2}{y}^{3}+2\,{x}^{3}+3\,xy}}.
\end{equation}
Let's compute $M_1$ and $N_1$ using the method sketched above:
\end{exem}

\noindent
{\bf Procedure:}

\begin{enumerate}

\item The candidates $M_c$, $N_c$ and ${\cal I}_c$ \footnote{We will not write them to save space. The names of the coefficients of $M_c$, $N_c$ and ${\cal I}_c$ are $m_i$, $n_j$ and $p_k$, respectively} are of degrees 6, 6 and 7, respectively (later on we will discuss these choices). We substitute them in the equation $\Delta - {\cal I}\, T_0=0$ and collect it in the variables $(x,y)$. We obtain a set of (linear) equations for the coefficients of the candidates.

\item We solve the set of (linear) equations for the coefficients of the candidates, obtaining the solution (we show only a few):
\begin{eqnarray}
\left\{ m_{0}=0,m_{1}={\frac {11}{15}}\,m_{26}-{\frac{7}{15}}\,n_{21}+\frac{2}{3}\,n_{14},m_{10}=9\,p_{21}+\frac{9}{5}\,n_{3}+{\frac {21}{10}}\,m_{13},
\right.
\nonumber \\ [2mm]
\left. m_{11}=-\frac{3}{2}\,p_{34}+\frac{3}{10}\,n_{18}-{\frac {9}{10}}\,m_{14},\cdots, p_{9}=-\frac{1}{2}\,p_{21}+\frac{1}{10}\,n_{3}-\frac{3}{10}\,m_{13} \right\}.
\nonumber 
\end{eqnarray}

\item We substitute the solution in the equation $D_1(\Delta) - {\cal I} \, (D_1(T_0) + T_0 \, T_1)=0$ and collect the equation in the variables $(x,y)$. Solving the set of equations for the remaining undetermined coefficients we obtain
\begin{eqnarray}
\left\{ m_{13}=2\,n_{3},m_{14}=0,m_{15}=0,m_{18}=0,m_{26}=0,n_{14}=0,n_{18}=0,n_{21}=0,\,\,
\right.
\nonumber \\ [2mm]
\left.  n_{26}=0,n_{3}=n_{3},n_{4}=0,p_{21}=-n_{3},p_{22}=0,p_{25}=0,p_{34}=0 \right\}, \qquad\qquad\qquad
\nonumber 
\end{eqnarray}

\item We substitute the solution above in the candidates $M_c$ and $N_c$, obtaining $M_1 = -y \left( -{y}^{5}+3\,x{y}^{2}-2 \right)$ and $N_1 =  -7\,x{y}^{5}+{x}^{2}{y}^{2}+5\,{y}^{3}+x$.

\end{enumerate}

\begin{obs}
\label{obsexamp2}
Some comments:

\begin{itemize}

\item The 1ODE$_0$ {\em (\ref{examp2eq1})} is the one presented in the example {\em \ref{examp1}}. Thus, the association of the method developed in this section with the `linear calculation' of the cofactors (that we've presented in section {\em \ref{lncc}}) generates a complete (semi) algorithm to determine Darboux polynomials. Besides, since the determination of the Darboux polynomials is the `hard part' (i.e., the computationally costly part) of the whole process, we can associate these two processes with the final (linear) part of the Darboux-Prelle-Singer algorithm (In the case where $I_0$ is an elementary function) or with the final (linear) part of the algorithm presented in \cite{Nosjcam2005} (In the case where $I_0$ is a non elementary Liouvillian function) and obtain a semi algorithm to compute Liouvillian first integrals.

\item The computation of $M_1$ and $N_1$ using the method described above spent less than 1s of CPU time and 32Mb of memory (running in a Maple platform).\footnote{For this kind of information see section \ref{perf}.}

\item The procedure, in the way it was built, applies only to rational 1ODEs such that $T_0 | \Delta$. However (as will be seen later), this condition is not as restrictive as it seems. 

\item The solution presented in item {\em 3} is not unique. There are another vector fields ${D}_{1j}$ such that the coefficients satisfy the equations {\em (\ref{reldeltat0})} and {\em (\ref{d1reldeltat0})}, but ${D}_{1j} (R_0) \neq 0$ (${D}_{1j} \in [D_1]$, i.e., ${D}_{1j} (R_{0j}) = 0,\,\,R_{0j} \in [R_0]$ -- see the remark \ref{classd1}). 
\end{itemize}
\end{obs}

\subsection{Some Improvements}
\label{si}

In the previous subsections we present two procedures that, put together, constitute a method to determine DPs of polynomial vector fields that present a Liouvillian first integral. In this subsection we present some improvements that can be used in the great majority of cases. We will show these improvements for the cases where the Liouvillian first integral $I_0$ is a function of one of the following types: 
\begin{itemize}
\item Liouvillian (non elementary Liouvillian function).
\item Elementary (non algebraic Darboux function).
\item Rational (non polynomial function).
\end{itemize}

\subsubsection{$I_0$ is a non elementary Liouvillian function}
\label{i0nel}

Remember that, for this case, the integrating factor $R_0$ is of the format
\begin{equation}
\label{r01}
R_0= {\rm e}^{Z_0}\,\prod_i {p_i}^{n_i},
\end{equation}
where $Z_0=A/B$ is a rational function ($A$ and $B$ are coprime polynomials), $p_i$ are irreductible polynomials, $n_i$ are rational numbers and the polynomials $p_i$ and $B$ are Darboux polynomials of the operator $D_0$. $R_0$ is (by definition) a first integral of $D_1$, then 
\begin{equation}
{R_0}_x = M_1\,R_1, \,\,\,\,\,\, {R_0}_y = -N_1\,R_1.
\end{equation}
Calculating the derivatives of $R_0$, we arrive at
\begin{eqnarray}
{R_0}_x &=& R_0\left(\frac{B\,\partial_x(A)-A\,\partial_x(B)}{B^2} + \sum_i n_i\,\frac{\partial_x(p_i)}{p_i}\right)=
\nonumber \\ [2mm]
&=& \frac{R_0}{lmc(B^2,\prod_i {p_i})} \left(B\,\partial_x(A)-A\,\partial_x(B) + \sum_i n_i\,\prod_{i \neq j} {p_j}^{n_j} \, \partial_x(p_i)\right),
\nonumber \\ [2mm]
{R_0}_y &=& R_0\left(\frac{B\,\partial_y(A)-A\,\partial_y(B)}{B^2} + \sum_i n_i\,\frac{\partial_y(p_i)}{p_i}\right)=
\nonumber \\ [2mm]
&=& \frac{R_0}{lmc(B^2,\prod_i {p_i})} \left(B\,\partial_y(A)-A\,\partial_y(B) + \sum_i n_i\,\prod_{i \neq j} {p_j}^{n_j} \, \partial_y(p_i)\right).
\nonumber 
\end{eqnarray}
So, in the general case 
\begin{equation}
\label{r0r1}
R_1= \frac{R_0}{lmc(B^2,\prod_i {p_i})}\,\,\,\,\, \Rightarrow\,\,\,\,\, {\cal I} = {lmc(B^2,\prod_i {p_i})}.
\end{equation}
Thus ${\cal I}$ is composed by a product of Darboux polynomials where the squared polynomials are the factors of $B$. Using this we can propose other ways to compute the Darboux polynomials: 

\begin{enumerate}

\item Instead of using the equation $D_1(\Delta) - {\cal I} \, (D_1(T_0) + T_0 \, T_1)=0$, we use $D_1({\cal I}) - {\cal I} \, T_1=0$. In addition, we do not need to use $M_1$ and $N_1$, since ${\cal I}$ is composed of the Darboux polynomials necessary (and sufficient) to construct the integrating factor $R_0$. 

\item We apply $D_0$ to the equation $\Delta - {\cal I}\,T_0$. We obtain: $D_0(\Delta) - {\cal I} \, (D_1(T_0) + T_0 \, Q_0)=0$, where $Q_0$ is the cofactor of ${\cal I}$ and so, $Q_0$ is a polynomial of maximum degree equal to ${\rm max}(\deg(M_0),\deg(N_0))$\footnote{Since $T_0$ is composed with the cofactors of the Darboux polynomials present in the integrating factor $R_0$ then if there is no cancellation of terms, the set $\{{\rm mon}(Q_0)\}$ of monomials of $Q_0$ is a subset of set $\{{\rm mon}(T_0)\}$.}. Therefore we can use $D_0(\Delta) - {\cal I} \, (D_1(T_0) + T_0 \, Q_0)=0$ instead of $D_1(\Delta) - {\cal I} \, (D_1(T_0) + T_0 \, T_1)=0$ and (as in the previous item) we do not need to use $M_1$ and $N_1$ for the same reason.

\item A possible subcase of the item 2 is to look for a vector field such that the inverse integrating factor ${\cal I}$ is $\prod_j p_j$, where the $\prod_j p_j$ is a product of all Darboux polynomials present in $D_0$. 
\end{enumerate}

\subsubsection{$I_0$ is a non algebraic elementary function}
\label{i0nre}

For this case, the integrating factor $R_0$ is of the format
\begin{equation}
\label{r02}
R_0= \prod_i {p_i}^{n_i},
\end{equation}
where $n_i$ are rational numbers and $p_i$ are irreductible Darboux polynomials of $D_0$. Following the reasoning used in the previous case we have
\begin{eqnarray}
{R_0}_x &=& R_0\left(\sum_i n_i\,\frac{\partial_x(p_i)}{p_i}\right)= \frac{R_0}{\prod_i {p_i}} \left(\sum_i n_i\,\prod_{i \neq j} {p_j}^{n_j} \, \partial_x(p_i)\right),
\nonumber \\ [2mm]
{R_0}_y &=& R_0\left(\sum_i n_i\,\frac{\partial_y(p_i)}{p_i}\right)= \frac{R_0}{\prod_i {p_i}} \left(\sum_i n_i\,\prod_{i \neq j} {p_j}^{n_j} \, \partial_y(p_i)\right).
\nonumber 
\end{eqnarray}
So, in the general case 
\begin{equation}
\label{r0r12}
R_1= \frac{R_0}{\prod_i {p_i}}\,\,\,\,\, \Rightarrow\,\,\,\,\, {\cal I} = \prod_i {p_i}.
\end{equation}
Therefore, the possible improvements are the same of the processes proposed on items 1 and 2 of subsection \ref{i0nel}.

\subsubsection{$I_0$ is a non polynomial rational function}
\label{i0npr}

The integrating factor $R_0$ is a rational function and, 
\begin{equation}
\label{r03}
R_0= \prod_i {p_i}^{n_i},
\end{equation}
where $n_i$ are integers and $p_i$ are irreductible Darboux polynomials of $D_0$. Following the reasoning used in the previous sections we have
\begin{equation}
\label{r0r13}
R_1= \frac{R_0}{\prod_i {p_i}}\,\,\,\,\, \Rightarrow\,\,\,\,\, {\cal I} = \prod_i {p_i},
\end{equation}
and we can use $D_1({\cal I}) - {\cal I} \, T_1=0$ or the second item of section \ref{i0nel} to compute the $p_i$ directly (like before). Besides, in this case we can use a heuristic method if $T_0$ is a polynomial with few terms. Since, $T_0$ is a weighted sum of cofactors then we can presume that there is a high probability of $D_0(\Pi_p)=\Pi_p\,\prod_k a_k\,{\rm mon}(T_0)_k$, where $\Pi_p = \prod_j {p_j}^{m_j}$, ${\rm mon}(T_0)$ is a set of the monomials of $T_0$, $a_k$ are constants to be determined and ${m_j}$ are integers.

\medskip

\noindent
Let's see one of these `improved procedures' in action:

\begin{exem}
\label{examp3}
Consider again the rational 1ODE$_0$ given by 
\begin{equation}
\label{examp2eq1}
\phi_0={\frac {-{y}^{7}+x{y}^{4}-{x}^{2}y+{y}^{2}}{2\,x{y}^{6}-7\,{x}^{2}{y}^{3}+2\,{x}^{3}+3\,xy}}.
\end{equation}
Let's compute the Darboux polynomials using the `improved method' sketched above:
\end{exem}

\noindent
{\bf Procedure:}

\begin{enumerate}

\item The candidates $M_c$, $N_c$ and ${\cal I}_c$ are of degree 6, 6 and 7, respectively. We substitute them in the equation $\Delta - {\cal I}\, T_0=0$ and collect it in the variables $(x,y)$. We obtain a set of (linear) equations for the coefficients of the candidates.

\item We solve the set of (linear) equations for the coefficients of the candidates, obtaining the solution.\footnote{So far everything exactly like the procedure described in section \ref{tsp}.}

\item We substitute the solution in the equation $D_1({\cal I}) - {\cal I} \, T_1=0$ and collect the equation in the variables $(x,y)$. Solving the set of equations for the remaining undetermined coefficients we obtain
\begin{eqnarray}
\left\{ m_{13}\!=0,m_{14}\!=0,m_{15}=\!-3\,p_{23},m_{18}=0,m_{19}=0,n_{14}=0,n_{18}=0,n_{19}=p_{23},\,\,
\right.
\nonumber \\ [2mm]
\left.  n_{22}=0,n_{3}=0,n_{4}=0,p_{21}=0,p_{22}=0,p_{23}=p_{23},p_{26}=0 \right\}, \qquad\qquad\qquad\qquad
\nonumber 
\end{eqnarray}

\item We substitute the solution above in ${\cal I}_c$, obtaining\footnote{The process uses $\approx$ 28Mb of memory and $\approx$ 1s of CPU time.}$$p_{23}\,y \left( x{y}^{2}-1 \right)  \left( -{y}^{3}+x \right).$$

\end{enumerate}

\medskip

\section{Methods and their performance}
\label{mp}

In this section we will present three methods to determine Liouvillian first integrals of planar polynomial vector fields and make a study of their efficiency. We will do this as follows:

\begin{enumerate}
\item In the first subsection we present the `step by step' of the two (main) methods and the improvements discussed in section 2. From now on we will call them: 
\begin{itemize}
\item $CoLin$ algorithm ({\em calculating Cofactors Linearly} -- the method that uses $M_1$ and $N_1$ to calculate the cofactors linearly). 
\item $SInGeR$ algorithm ({\em Some Integrating factors Generated Rapidly} -- the method that computes the Darboux polynomials directly).
\item $ImpA$ algorithm ({\em Improved Algorithms} -- the improvements presented on section \ref{si}).
\end{itemize}

\item In subsection \ref{perf} we make a comparison of the three methods (and the MUC) and study their performance.
\end{enumerate}

\subsection{Three possible methods}
\label{tpm}

In the following three subsections we will translate the methods presented in section 2 into three algorithms (semi).

\subsubsection{The $CoLin$ algorithm}
\label{colin}

\begin{algor}[$CoLin$]$\!$: This algorithm is based on the method described in the sections \ref{lncc} and \ref{o1odea}.
\
\vspace{3mm}

{\bf Steps:}
\begin{enumerate}

\item Choose $d_g$ (a positive integer) for the degree of the candidates $M_c$ and $N_c$.

\item Choose $d_{\cal I} = d_g + 1$ for the degree of the candidate ${\cal I}_c$.

\item Construct three polynomials $M_c$, $N_c$ and ${\cal I}_c$ of degrees $d_g$, $d_g$ and $d_{\cal I}$, respectively, with undetermined coefficients. 

\item Substitute then in the equation $E_1\!: \Delta - {\cal I}\, T_0=0$.

\item  Collect the equation $E_1$ in the variables $(x,y)$ obtaining a set of (linear) equations $S_{E_1}$ for the coefficients of the polynomials $M_c$, $N_c$ and ${\cal I}_c$. 

\item Solve $S_{E_1}$ to the undetermined coefficients. 

\item Substitute the solution of $S_{E_1}\!\!$ in the equation $E_2\!:\! D_1(\Delta) - {\cal I} \, (D_1(T_0) + T_0 \, T_1)\!=\!0$.

\item Collect $E_2$ in the variables $(x,y)$ obtaining a set of equations $S_{E_2}$ for the remaining undetermined coefficients of the polynomials $M_c$, $N_c$ and ${\cal I}_c$. 

\item Solve $S_{E_2}$ for the remaining undetermined coefficients. If no solution is found then go to item 2.

\item Construct an empty list $L_{S_3}=[\,\,\,]$ .

\item For each of the solutions of $S_{E_2}$ do:

\begin{enumerate} 

\item Substitute the solution of $S_{E_2}$ in $M_c$ and $N_c$ to obtain $M_1$, $N_1$.

\item Construct the operators $D_0$ and $D_1$ and the functions $F_0$ and $F_1$.

\item Construct two polynomials $Q_0$ and $Q_1$ (the cofactor candidates) of degrees $\max(deg_{M_0},deg_{N_0})$ and $\max(deg_{M_1},deg_{N_1})$.

\item Substitute $Q_0$ and $Q_1$ in the equation $E_3: D_0(Q_1)-D_1(Q_0)-Q_0\,F_0-Q_1\,F_1=0$.

\item Collect the numerator of $E_3$ in the variables $(x,y)$, obtaining a set of (linear) equations $S_{E_3}$ for the coefficients of $Q_0$ and $Q_1$.

\item Solve $S_{E_3}$ for the coefficients of $Q_0$ and $Q_1$. 

\item If the solution of $S_{E_3}$ is non trivial then add it to $L_{S_3}$.

\end{enumerate}

\item If $L_{S_3}=[\,\,\,]$ then go to item 2.

\item Construct an empty list $L_{S_4}=[\,\,\,]$ .

\item For each of the solutions in $L_{S_3}$ do:

\begin{enumerate} 

\item Choose $d_p = c$ ($c$ is a positive integer) for the degree of the Darboux polynomial candidate $P_c$.

\item Construct a polynomial $P_c$ of degree $d_p$. 

\item Substitute $P_c$ in the equation $E_4: D_1(P_c)-Q_1\,P_c=0$.

\item Collect $E_4$ in the variables $(x,y)$, obtaining a set of equations $S_{E_4}$ for the coefficients of $P_c$ and the remaining undetermined coefficients of the polynomial $Q_1$. 

\item Solve $S_{E_4}$ for the undetermined coefficients. 

\item Add each non trivial solution of $S_{E_4}$ to $L_{S_4}$. Each one of them corresponds to a Darboux polynomial.

\end{enumerate}

\item If $L_{S_4}=[\,\,\,]$ then go to item 2.

\item Use the Darboux polynomials found in the final part of the DPS algorithm (resp. in the final part of the algorithm presented in \cite{Nosjcam2005}) to determime an integrating factor $R_0$.

\item  If no integrating factor $R_0$ can be constructed then go to item 2.

\item Construct the first integral $I_0$.

\end{enumerate}

\end{algor}

\begin{obs}
Some comments:

\begin{itemize}

\item A closer examination of the steps shows us that the procedure may never end, i.e., more formally, $CoLin$ is a semi algorithm.

\item The choice for $d_{\cal I}$ comes from the fact that ${\cal I} = \Delta/T_0$.

\item Some of the steps described above are obviously much more complicated than others. Some involve very complex algorithms in themselves, for example, the algorithm that solves nonlinear polynomial systems.

\item Basically, the $CoLin$ algorithm divides the (quadratic) problem of solving the system resulting from the equation $D_0(p)-p\,q_0=0$ into two linear problems and two quadratic problems. Each of the quadratic problems involves solving a particular MUC. The expectation is that, in the vast majority of cases, the time and memory spent by the two MUCs will be much less than that spent by the original MUC.

\item In \cite{FeGa} Ferragut and Gasull envisioned a procedure to simplify (quite) this type of MUC. We are not discussing in this work the (actual) possibility of simplifying the two quadratic problems using the FG method.

\end{itemize}

\end{obs}

\subsubsection{The $SInGeR$ algorithm}
\label{singe}

\begin{algor}[$SInGeR$]$\!$: This algorithm is based on the improvement described in the item 1 of section \ref{i0nel}.
\
\vspace{3mm}

{\bf Steps:}
\begin{enumerate}

\item Choose $d_g$ (a positive integer) for the degree of the candidates $M_c$ and $N_c$.

\item Choose $d_{\cal I} = d_g + 1$ for the degree of the candidate ${\cal I}_c$.

\item Construct three polynomials $M_c$, $N_c$ and ${\cal I}_c$ of degrees $d_g$, $d_g$ and $d_{\cal I}$, respectively, with undetermined coefficients. 

\item Substitute then in the equation $E_1\!: \Delta - {\cal I}\, T_0=0$.

\item  Collect the equation $E_1$ in the variables $(x,y)$ obtaining a set of (linear) equations $S_{E_1}$ for the coefficients of the polynomials $M_c$, $N_c$ and ${\cal I}_c$. 

\item Solve $S_{E_1}$ to the undetermined coefficients. 

\item Substitute the solution of $S_{E_1}\!\!$ in the equation $E_2: D_1({\cal I}) - {\cal I} \, T_1=0$.

\item Collect $E_2$ in the variables $(x,y)$ obtaining a set of equations $S_{E_2}$ for the remaining undetermined coefficients of the polynomials $M_c$, $N_c$ and ${\cal I}_c$. 

\item Solve $S_{E_2}$ for the remaining undetermined coefficients. If no solution is found then go to item 2.

\item Construct an empty list $L_{S_3}=[\,\,\,]$.

\item For each of the solutions of $S_{E_2}$ do:

\begin{enumerate} 

\item Substitute the solution of $S_{E_2}$ in ${\cal I}_c$ to obtain a product of Darboux polynomials.

\item Add the Darboux polynomials to $L_{S_3}$.

\end{enumerate}

\item If $L_{S_3}=[\,\,\,]$ then go to item 2.

\item Use the Darboux polynomials found in the final part of the DPS algorithm (resp. in the final part of the algorithm presented in \cite{Nosjcam2005}) to determime an integrating factor $R_0$.

\item  If no integrating factor $R_0$ can be constructed then go to item 2.

\item Construct the first integral $I_0$.

\end{enumerate}

\end{algor}

\begin{obs}
Some comments:

\begin{itemize}

\item Like the $CoLin$, the $SInGeR$ procedure may never end, i.e., $SInGeR$ is also a semi algorithm.

\item Until step six, the procedures $CoLin$ and $SInGeR$ are the same. Therefore, whenever solving the $E_2$ equation of the $SInGeR$ algorithm is faster (or less expensive) than all the other steps of the $CoLin$ algorithm, the $SInGeR$ procedure will be more efficient.

\item The $SInGeR$ algorithm is extremely efficient for finding non rational elementary first integrals (see section \ref{perf}).
\end{itemize}

\end{obs}

\subsubsection{The $ImpA$ algorithm}
\label{impa}

\begin{algor}[$ImpA$]$\!$: \footnote{In reality, this algorithm is the collection of the improvements described in the items 2 and 3 described in the section \ref{i0nel}.}
\
\vspace{3mm}

{\bf Steps of $ImpA$:}
\begin{enumerate}

\item Choose $d_g $ (a positive integer) for the degree of the candidates $M_c$ and $N_c$.

\item Choose $d_{\cal I} = d_g + 1$ for the degree of the candidate ${\cal I}_c$.

\item Construct three polynomials $M_c$, $N_c$ and ${\cal I}_c$ of degrees $d_g$, $d_g$ and $d_{\cal I}$, respectively, with undetermined coefficients. 

\item Substitute then in the equation $E_1\!: \Delta - {\cal I}\, T_0=0$.

\item  Collect the equation $E_1$ in the variables $(x,y)$ obtaining a set of (linear) equations $S_{E_1}$ for the coefficients of the polynomials $M_c$, $N_c$ and ${\cal I}_c$. 

\item Solve $S_{E_1}$ to the undetermined coefficients. 

\item Construct a polynomial $Q_0$ of degree $d_{q_0}={\rm max}({\rm deg}(M_0),{\rm deg}(N_0))$ with undetermined coefficients. 

\item Substitute the solution of $S_{E_1}\!\!$ in the equation $\!E_2\!:\! D_0(\Delta)\! - \!{\cal I} (D_0(T_0) + T_0 \, Q_0)\!=\!0$.

\item Collect $E_2$ in the variables $(x,y)$ obtaining a set of equations $S_{E_2}$ for the remaining undetermined coefficients of$M_c$, $N_c$, ${\cal I}_c$ and the coefficients of $Q_0$. 

\item Solve $S_{E_2}$ for the remaining undetermined coefficients. If no solution is found then go to item 2.

\item Construct an empty list $L_{S_3}=[\,\,\,]$ .

\item For each of the solutions of $S_{E_2}$ do:

\begin{enumerate} 

\item Substitute the solution of $S_{E_2}$ in ${\cal I}_c$.

\item If the result is non trivial then add it to $L_{S_3}$.

\end{enumerate}

\item If $L_{S_3}=[\,\,\,]$ then go to item 2.

\item Use the Darboux polynomials found in the final part of the DPS algorithm (resp. in the final part of the algorithm presented in \cite{Nosjcam2005}) to determime an integrating factor $R_0$.

\item  If no integrating factor $R_0$ can be constructed then go to item 2.

\item Construct the first integral $I_0$.

\end{enumerate}

\end{algor}

\begin{obs}
Some comments:
\begin{itemize}

\item $ImpA$ is (like $CoLin$ and $SInGeR$) a semi algorithm.

\item If the time spent (or the use of memory) is excessive we can test a possible heuristic: To construct the polynomial $Q_0$ use $\{{\rm mon}(T_0)\}$.

\item The difference between the implementation of the algorithms described in items 2 and 3 of section \ref{i0nel} is just a matter of choosing different degrees.
\end{itemize}
\end{obs}

\subsection{Performance}
\label{perf}

In this section we will test the algorithms by comparing their memory usage and CPU time spent for a set of vector fields that have (mostly) high-degree Darboux polynomials in the integrating factors. In this scenario we will perform the test on a notebook (intel I5 processor - 4GB) running a maple platform.

\subsubsection{Example 1}
\label{exa1}
\begin{equation}
\phi_0={\frac {\!\!\!\!-\!16\,x{y}^{6}\!+\!32\,{x}^{2}{y}^{4}\!-\!16\,{x}^{3}{y}^{2}\!+\!12\,{y}^{5
}\!-\!16\,{x}^{2}{y}^{2}\!-24x{y}^{3}\!+\!12{x}^{2}y+4\,{y}^{3}+20\,xy-9}{16
\,{x}^{2}{y}^{5}\!-\!32\,{x}^{3}{y}^{3}\!+\!16\,{x}^{4}y\!-\!32{x}^{2}{y}^{3}\!-\!12x{y}^{4}\!+\!24{x}^{2}{y}^{2}\!-\!12{x}^{3}\!+\!44x{y}^{2}\!+\!4{x}^{2}\!-\!18y}} \nonumber
\end{equation}
\begin{table}[h]
{\begin{center} {\footnotesize
\begin{tabular}
{|c|c|c|c|}
\hline
Method & Memory & Time & DPs \\
\hline
 $CoLin$ & 35MB & 1.19s  &  $4\,xy-3,\, -{y}^{2}+x$ \\
\hline
 $SInGeR$ & 28MB & 0.90s  &  $4\,xy-3,\, -{y}^{2}+x$ \\
\hline
 $ImpA$ & 4MB & 0.09s  &  $4\,xy-3,\, -{y}^{2}+x$ \\
\hline
 $MUC$ & 46MB & 4.12s  &  $4\,xy-3,\, -{y}^{2}+x$ \\
\hline
\end{tabular} }
\end{center}}
\label{tex1}
\end{table}

\medskip

\subsubsection{Example 2}
\label{exa2}
\begin{eqnarray}
\phi_0&=& \!\!\left(72{x}^{5}{y}^{4}\!-\!48{x}^{5}{y}^{3}+36\,{x}^{3}{y}^{5}+8\,{x
}^{5}{y}^{2}-24\,{x}^{3}{y}^{4}+36{x}^{4}{y}^{2}+4{x}^{3}{y}^{3}-12{x}^{4}y+ \right.
\nonumber \\ [2mm]
&& 
\left. +12{x}^{2}{y}^{3}-16{x}^{2}{y}^{2}-3{y}^{4}+8{x}^{
3}+4\,{x}^{2}y+{y}^{3}+4\,xy-4\,x \right) / \left(-18\,{x}^{4}{y}^{4}+ \right.
\nonumber \\ [2mm]
&& 
 +12{x}^{4}{y}^{3}\!-\!9{x}^{2}{y}^{5}\!+\!24{x}^{5}y-2{x}^{4}{y}^{2}\!+\!6{x}^{2}{y}^{4}\!
-\!4{x}^{5}+12\,{x}^{3}{y}^{2}-{x}^{2}{y}^{3}+4\,x{y}^{2}+
\nonumber \\ [2mm]
&& 
\left. -2\,{x}^{2}-xy-y+1 \right) \nonumber
\end{eqnarray}
\begin{table}[h]
{\begin{center} {\footnotesize
\begin{tabular}
{|c|c|c|c|}
\hline
Method & Memory & Time & DPs \\
\hline
 $CoLin$ & 350MB & 60s  &  Negative \\
\hline
 $SInGeR$ & 16MB & 0.95s  &  $3\,x\,y^2-x\,y+1,\, 2\,x^{2}+y$ \\
\hline
 $ImpA$ & 34MB & 1.17s  &  $3\,x\,y^2-x\,y+1,\, 2\,x^{2}+y$ \\
\hline
 $MUC$ & 350MB & 120s  &  Negative \\
\hline
\end{tabular} }
\end{center}}
\label{tex1}
\end{table}

\medskip

\subsubsection{Example 3}
\label{exa3}
\begin{equation}
\phi_0={\frac {-4x{y}^{16}+8{x}^{2}{y}^{12}+6{y}^{12}-4{x}^{3}{y}^{8}
-4{x}^{2}{y}^{8}-12x{y}^{8}+2{y}^{8}+6{x}^{2}{y}^{4}+10x{y}^
{4}-9}{\!\!4(4{x}^{2}{y}^{15}\!-\!8{x}^{3}{y}^{11}\!-\!4{x}^{2}{y}^{11}\!-\!6x{y}^{11}\!+\!4{x}^{4}{y}^{7}\!+\!12{x}^{2}{y}^{7}\!+\!10x{y}^{7}\!-\!6{x}^{3}{y}^{3}\!+\!2{x}^{2}{y}^{3}\!-\!9{y}^{3})}} \nonumber
\end{equation}
\begin{table}[h]
{\begin{center} {\footnotesize
\begin{tabular}
{|c|c|c|c|}
\hline
Method & Memory & Time & DPs \\
\hline
 $CoLin$ & 222MB & 19.34s  &  $2\,xy^4-3,\, -{y}^{4}+x$ \\
\hline
 $SInGeR$ & 307MB & 120s  &  Negative \\
\hline
 $ImpA$ & 16MB & 0.93s  &  $2\,xy^4-3,\, -{y}^{4}+x$ \\
\hline
 $MUC$ & 210MB & 120s  &  Negative \\
\hline
\end{tabular} }
\end{center}}
\label{tex1}
\end{table}

\medskip

\subsubsection{Example 4}
\label{exa4}
\begin{equation}
\phi_0={\frac {-2\,{x}^{3}{y}^{19}+2\,{x}^{5}{y}^{14}-{y}^{17}-2\,{x}^{2}{y}^
{12}+3\,{x}^{4}{y}^{7}+2\,{x}^{2}{y}^{7}-2\,x{y}^{5}+2\,{x}^{3}+2\,x}{
{y}^{4} \left( -5\,{x}^{2}{y}^{19}+5\,{x}^{4}{y}^{14}-3\,x{y}^{12}-4\,
{x}^{3}{y}^{7}+5\,x{y}^{7}+7\,{x}^{5}{y}^{2}-5\,{y}^{5}+5\,{x}^{2}+5
 \right) }} \nonumber
\end{equation}
\begin{table}[h]
{\begin{center} {\footnotesize
\begin{tabular}
{|c|c|c|c|}
\hline
Method & Memory & Time & DPs \\
\hline
 $CoLin$ & 210MB & 5.10s  &  $x\,y^7+1,\,y^5-x^2$ \\
\hline
 $SInGeR$ & 16MB & 0.84s  &  $x\,y^7+1,\,y^5-x^2$ \\
\hline
 $ImpA$ & 30MB & 1.21s  &   $x\,y^7+1,\,y^5-x^2$ \\
\hline
 $MUC$ & 350MB & 80s  &  Negative \\
\hline
\end{tabular} }
\end{center}}
\label{tex1}
\end{table}

\medskip

\subsubsection{Example 5}
\label{exa5}
\begin{equation}
\phi_0={\frac {-y \left( 4\,{y}^{12}-{y}^{9}-20\,x{y}^{6}+3\,{y}^{5}+36
\,{x}^{2} \right) }{2(-2\,{y}^{15}+2\,x{y}^{12}+15\,x{y}^{9}-24\,{x}^{2}
{y}^{6}-9\,x{y}^{5}-18\,{x}^{2}{y}^{3}+18\,{x}^{3})}} \nonumber
\end{equation}
\begin{table}[h]
{\begin{center} {\footnotesize
\begin{tabular}
{|c|c|c|c|}
\hline
Method & Memory & Time & DPs \\
\hline
 $CoLin$ & 16MB & 1.03s  &  $y^4-4\,xy-3,\,y^6-3x$ \\
\hline
 $SInGeR$ & 46MB & 3.40s  &  $y^4-4\,xy-3,\,y^6-3x$ \\
\hline
 $ImpA$ & 16MB & 0.28s  &   $y^4-4\,xy-3,\,y^6-3x$ \\
\hline
 $MUC$ & 350MB & 60s  &  Negative \\
\hline
\end{tabular} }
\end{center}}
\label{tex5}
\end{table}

\bigskip
\bigskip

\subsubsection{Example 6}
\label{exa6}
\begin{equation}
\phi_0={\frac {6\,{x}^{7}{y}^{10}+{x}^{14}{y}^{2}+{y}^{13}-2\,{x}^{7}{y}^{5}-
6\,{x}^{8}y+{y}^{9}-2\,{x}^{7}y+{y}^{8}-x{y}^{4}+2\,{y}^{4}-x+1}{9{x
}^{14}{y}^{10}\!-\!18{x}^{7}{y}^{13}\!-\!{x}^{8}{y}^{9}\!-\!18{x}^{7}{y}^{9}\!+\!9{y}^{16}\!+\!4x{y}^{12}+18\,{y}^{12}+{x}^{9}+9\,{y}^{8}-4\,{x}^{2}{y}^{3}}} \nonumber
\end{equation}
\begin{table}[h]
{\begin{center} {\footnotesize
\begin{tabular}
{|c|c|c|c|}
\hline
Method & Memory & Time & DPs \\
\hline
 $CoLin$ & 213MB & 5.31s  &  $x^7y-y^4-1,\,-y^9+x$ \\
\hline
 $SInGeR$ & 30MB & 1.51s  &  $x^7y-y^4-1,\,-y^9+x$ \\
\hline
 $ImpA$ & 30MB & 1.73s  &   $x^7y-y^4-1,\,-y^9+x$ \\
\hline
 $MUC$ & 186MB & 120s  &  Negative \\
\hline
\end{tabular} }
\end{center}}
\label{tex6}
\end{table}

\bigskip
\bigskip

\subsubsection{Example 7}
\label{exa7}
\begin{equation}
\phi_0={\frac {6\,{x}^{7}{y}^{10}+{x}^{14}{y}^{2}+{y}^{13}-2\,{x}^{7}{y}^{5}-
6\,{x}^{8}y+{y}^{9}-2\,{x}^{7}y+{y}^{8}-x{y}^{4}+2\,{y}^{4}-x+1}{9{x
}^{14}{y}^{10}\!-\!18{x}^{7}{y}^{13}\!-\!{x}^{8}{y}^{9}\!-\!18{x}^{7}{y}^{9}\!+\!9{y}^{16}\!+\!4x{y}^{12}+18\,{y}^{12}+{x}^{9}+9\,{y}^{8}-4\,{x}^{2}{y}^{3}}} \nonumber
\end{equation}
\begin{table}[h]
{\begin{center} {\footnotesize
\begin{tabular}
{|c|c|c|c|}
\hline
Method & Memory & Time & DPs \\
\hline
 $CoLin$ & 205MB & 5.31s  &  $x^4y^2-2\,x^3y+x^2+3,\,x$ \\
\hline
 $SInGeR$ & 31MB & 1.10s  &  $x^4y^2-2\,x^3y+x^2+3,\,x$ \\
\hline
 $ImpA$ & 4MB & 0.06s  &   $x^4y^2-2\,x^3y+x^2+3,\,x$ \\
\hline
 $MUC$ & 500MB & 120s  &  Negative \\
\hline
\end{tabular} }
\end{center}}
\label{tex7}
\end{table}

\bigskip
\bigskip

\subsubsection{Example 8}
\label{exa8}
\begin{equation}
\phi_0={\frac {{x}^{6}\!-\!2{x}^{5}y\!+\!3{x}^{4}y\!-\!4{x}^{3}{y}^{2}\!-\!3{x}^{4}\!+\!
4{x}^{3}y\!-\!3{x}^{2}{y}^{2}\!+\!2x{y}^{3}\!-\!{y}^{3}\!+\!3{x}^{2}\!-\!2xy\!+\!{y}
^{2}\!+\!y\!-\!1}{-({x}^{6}-{x}^{5}+2\,{x}^{4}y-{x}^{4}+2\,{x}^{3}y-{x}^{2}{y}^{
2}+x{y}^{2}-{x}^{2}-2\,xy+{y}^{2}+x-2\,y+1)}} \nonumber
\end{equation}
\begin{table}[h]
{\begin{center} {\footnotesize
\begin{tabular}
{|c|c|c|c|}
\hline
Method & Memory & Time & DPs \\
\hline
 $CoLin$ & 4MB & 0.09s  &  $x^4+y^2-1$ \\
\hline
 $SInGeR$ & 4MB & 0.09s  &  $x^4+y^2-1$ \\
\hline
 $ImpA$ & 4MB & 0.09s  &   $x^4+y^2-1$ \\
\hline
 $MUC$ & 350MB & 60s  &  Negative \\
\hline
\end{tabular} }
\end{center}}
\label{tex8}
\end{table}

\newpage

\subsubsection{Example 9}
\label{exa9}
\begin{equation}
\phi_0={\frac {-2x \left( -16\,{x}^{6}{y}^{9}+8{x}^{14}\!-\!18{x}^{4}{y}^{
10}-2\,{y}^{13}\!+\!10{x}^{8}{y}^{4}\!-\!2{x}^{2}{y}^{10}\!-\!2{x}^{10}y-3\,
{y}^{11} \right) }{18\,{x}^{8}{y}^{8}+20\,{x}^{6}{y}^{9}+6\,{x}^{2}{y}
^{12}-24\,{x}^{10}{y}^{3}+6\,{x}^{4}{y}^{9}+4\,{y}^{13}+3\,{x}^{12}+7
\,{x}^{2}{y}^{10}}} \nonumber
\end{equation}
\begin{table}[h]
{\begin{center} {\footnotesize
\begin{tabular}
{|c|c|c|c|}
\hline
Method & Memory & Time & DPs \\
\hline
 $CoLin$ & 4MB & 0.09s  &  $2\,{x}^{6}-2\,{y}^{4}+{x}^{2}y$ \\
\hline
 $SInGeR$ & 4MB & 0.06s  &  $2\,{x}^{6}-2\,{y}^{4}+{x}^{2}y$ \\
\hline
 $ImpA$ & 4MB & 0.07s  &   $2\,{x}^{6}-2\,{y}^{4}+{x}^{2}y$ \\
\hline
 $MUC$ & 235MB & 120s  &  Negative \\
\hline
\end{tabular} }
\end{center}}
\label{tex9}
\end{table}

\medskip

\subsubsection{Example 10}
\label{exa10}
\begin{equation}
\phi_0=-{\frac {-3\,{y}^{2}+x+3\,y}{x \left( 8\,y-9 \right) }} \nonumber
\end{equation}
\begin{table}[h]
{\begin{center} {\footnotesize
\begin{tabular}
{|c|c|c|c|}
\hline
Method & Memory & Time & DPs \\
\hline
 $CoLin$ & 223MB & 19.21s  &  ${y}^{4}\!+\!2{y}^{2}x\!+\!{x}^{2}\!-\!6yx,\,4{y}^{4}\!+\!8{y}^{2}x\!-\!4{y}^{3}\!+\!4{x}^{2}\!-\!36yx\!+\!27x ,\,x$ \\
\hline
 $SInGeR$ & 211MB & 15s  &  ${y}^{4}\!+\!2{y}^{2}x\!+\!{x}^{2}\!-\!6yx,\,4{y}^{4}\!+\!8{y}^{2}x\!-\!4{y}^{3}\!+\!4{x}^{2}\!-\!36yx\!+\!27x ,\,x$ \\
\hline
 $ImpA$ & 28MB & 0.98s  &   ${y}^{4}\!+\!2{y}^{2}x\!+\!{x}^{2}\!-\!6yx,\,4{y}^{4}\!+\!8{y}^{2}x\!-\!4{y}^{3}\!+\!4{x}^{2}\!-\!36yx\!+\!27x ,\,x$ \\
\hline
 $MUC$ & 54MB & 4.37s  &  ${y}^{4}\!+\!2{y}^{2}x\!+\!{x}^{2}\!-\!6yx,\,4{y}^{4}\!+\!8{y}^{2}x\!-\!4{y}^{3}\!+\!4{x}^{2}\!-\!36yx\!+\!27x ,\,x$ \\
\hline
\end{tabular} }
\end{center}}
\label{tex10}
\end{table}

\medskip

\subsubsection{Some comments}
\label{sc}

\begin{itemize}

\item As we did not have an extensive set of vector fields whose integrating factors had high degree Darboux polynomials, we decided to build some examples on our own. Because of this, the choice of examples followed some rather unusual criteria:
\begin{itemize}
\item In first place, it is very difficult to have a notion (a priori) of the `algorithmic difficulty'\footnote{Since we are working with different algorithms and we cannot favor any.} involved in the searching of DPs for a given vector field at the time we created it. The only clue is the degree of the Darboux polynomials involved and the degree of (and number of terms in) the $ M_0 $ and $ N_0 $ polynomials that form the vector field. So, we decided that the 1ODEs (associated with the vector fields $ D_0 $) would have only one line (the only exception is example 2).
\item The vast majority of vector fields with non-elementary Liouvillian first integrals and high degree DPs do not fit in half a page. Thus, the vector fields with non-elementary Liouvillian first integrals have DPs of lower degrees or with few monomials (examples 1, 2, 3 and 4).
\item The numerical coefficients of the monomials (for a reason similar to that presented in the previous item) are very simple.
\item The vector fields presenting elementary (non algebraic) first integrals (examples 5, 6) follow the same principle, i.e., the respective first integrals are not especially creative except for the example 7 -- example 181 of Kamke's book \cite{Kam}.
\item Finally, for the vector fields that present rational first integrals, we use three examples presented in A. Bostan {\it et al} \cite{BoChClWe}.
\end{itemize}

\item The CPU time spent implies only the time to find the DPs and not the time to find the first integrals. 

\item In the $CoLin$ algorithm the `linear part' (the calculation of the cofactors) is not included because it spends almost nothing. Furthermore, as we have seen, the first part is enough to find the DPs.

\item When we load the basic packages to start the Maple session, it begins by using 4MB and 0.06 sec. In this sense, these numbers can be translated as: `almost nothing'.

\item The times spent by our algorithms for the examples 8, 9 and 10, chosen from the paper \cite{BoChClWe}, cannot be compared to the times shown in the paper because we only count the computation times to the correct values of the degrees of the polynomials involved (i.e., the choices that do not result positively have not been computed).

\item The idea of vector fields that `share' the Darboux polynomials seems to be very fruitful and there is still much to be studied and improved. For example, we can look for vector fields that share not all of the Darboux polynomials present in the integral factor.

\item We consider only the case ({\it iii}) of theorem \ref{delit0} (because, in this case, the inverse integrating factor of $D_1$ is a polynomial). Nevertheless, this seems to be the general case at least from a practical point of view (that is, this situation seems to occur in almost all cases).

\end{itemize}

\section{Conclusion}
\label{conclu}

In this work we presented three efficient semi algorithms to compute Liouvillian first integrals for polynomial plane vector fields. In reality, the algorithms compute the Darboux polynomials (which are the building blocks for the construction of an integrating factor) of the vector fields (the complicated part of the process). The basic idea was to separate the Darboux polynomial calculation process into smaller procedures that could be finished in a `reasonable time'. We did it by computing an associated polynomial vector field that share (with the vector field in question) the Darboux polynomials. We also presented a brief comparison among them (we also compare with the naive method - MUC).

The basic idea can possibly be extended to polynomial vector fields in $\R^3$ that present Liouvillian first integrals. We intend to do this in a future work.


\end{document}